\documentclass[twocolumn,appendixfloats,tighten]{aastex6}
\usepackage{newtxtext,newtxmath}
\usepackage[T1]{fontenc}
\usepackage{ae,aecompl}

\usepackage{amsmath}	% Advanced maths commands
\usepackage{amssymb}	% Extra maths symbols

\usepackage{ctable}
\usepackage{url}
\usepackage{xspace}
\usepackage[normalem]{ulem}
\usepackage{hyperref}
\usepackage[all]{hypcap}
\usepackage{graphicx}
\usepackage{color}
\newcommand{\be}{\begin{equation}}
\newcommand{\ee}{\end{equation}}

\def\h2{${\rm\,H_2}$}

\def\gsim{ \lower .75ex \hbox{$\sim$} \llap{\raise .27ex \hbox{$>$}} }
\def\lsim{ \lower .75ex \hbox{$\sim$} \llap{\raise .27ex \hbox{$<$}} }

\usepackage{color}

\slugcomment{Accepted by ApJL, September 17, 2018}

\begin{document}

\title{Modeling Star Formation as a Markov Process in  a Supersonic Gravoturbulent Medium}
\author{Evan Scannapieco and Mohammadtaher Safarzadeh}

\affiliation{Arizona State University School of Earth and Space Exploration, P.O. Box 871404, Tempe, AZ 
85287, USA}

\begin{abstract}

Molecular clouds exhibit lognormal probability density functions (PDF) of mass densities, which are thought to arise as a consequence of isothermal, supersonic turbulence. Star formation is then widely assumed to occur in  perturbations in which gravitational collapse is faster than the rate of change due to turbulent motions. Here we use direct numerical simulations  to measure this rate as a function of density for a range of turbulent Mach numbers, and show it is faster at high densities than at low densities. Furthermore, we show that both the density PDF and  rate of change arise naturally in a simple model of turbulence as a continuous Markov process. The one-dimensional Langevin equation that describes this evolution depends on only two parameters,  yet it captures the full evolution seen in direct three-dimensional simulations.  If it is modified to include gravity, the Langevin equation also reproduces the rate of material collapsing to high densities seen in turbulent simulations including self-gravity.  When generalized to include both temperature and density, similar analyses are likely applicable throughout astrophysics.

\end{abstract}


\section{Introduction}

Many astrophysical systems exist in pseudo steady-states, in which the global properties are roughly constant over many dynamical times.  The interstellar medium, for example, can be thought of in this manner, with material moving between different phases whose mass fractions and properties remain roughly constant \citep{Cox05}.   Similarly, molecular clouds are well modeled as supersonic, isothermal systems that are in near virial equillibrium and driven for a sufficiently long times to exhibit a well-developed turbulent cascade \citep{Federrath:2010ef}.

The densities and temperatures of any such system will be distributed according to a probability distribution function (PDF) that varies slowly with time, even though individual parcels of material are constantly changing. In the case of isothermal turbulence, the mass-weighted PDF is given as
\be
P_M(s) \approx  \frac{1}{\sqrt{2 \pi \sigma_s^2}} {\rm exp} \left[ - \frac{(s - s_0)^2} {2 \sigma_s^{2}} \right],
\label{eq:PDF}
\ee
where $s \equiv \ln(\rho/\rho_0),$ $\rho_0$ is the mean density, and the variance $\sigma_s^2 = 2 s_0 $
as required by mass conservation \citep{VazquezSemadeni94,Padoan97,Federrath:2010ef}.  Direct numerical simulations 
find that $\sigma_s^2 = \ln(1 + b^2 M^2)$ where $M = \sigma_v/c_s$ is the ratio of the mean velocity dispersion $\sigma_v$ to the sound speed $c_s$, and $b\approx$ 1 and $1/3$ for compressive and solenoidal forcing, respectively  \citep{Padoan97,Ostriker2001,Price2011}.  

While the structure and lifetime of dense regions has been studied \citep{Falceta2011,Robertson2018}, the overall manner in which material moves within the turbulent PDF has never been directly measured. Nevertheless, it is this motion that controls which parcels of gas will form stars, as it determines which will collapse due to gravity before they are reshuffled to lower densities  \citep{Krumholz05,Padoan11,Hennebelle11,Federrath2012,Hopkins2013}, see however  \citep{Elmegreen2000,Murray2011}.  

Here we measure this overall evolution for the first time, show that it can be described by two simple functions, and show how these results can be extended to reproduce the results of gravoturbulent simulations of star formation.  The structure of this work is as follows:  In \S2 we describe our turbulent simulations and measurements of the evolution of the PDF.  In \S3 we apply these results to develop a Markov model of the evolution of the medium.  In \S4 we extend this model to  study star-formation in a steady-state gravoturbulent medium and in \S5 we show how similar models can be constructed to describe other scenarios of star formation as well as other astrophysical systems.

\section{Simulations}

To study the evolution of material in isothermal turbulence, we carried out direct numerical simulations over a range of Mach numbers using FLASH \citep{Fryxell:2000em} version 4.2.1. We solved the hydrodynamics equations using an unsplit solver with third-order reconstruction \citep{Lee2013}, and employed a hybrid Riemann solver, which uses both an extremely accurate but somewhat fragile Harten-Lax-van Leer-Contact (HLLC) solver \citep{Toro:1994gu} and a more robust, but more diffusive Harten Lax and van Leer (HLL) solver \citep{Einfeldt:1991kd}

Each simulation was carried out in a $512^3$ periodic box of size $L_{\rm box}$, within which turbulence was continuously driven \citep{Eswaran:1988vc}, by solenoidal modes (i.e.\ $\nabla \cdot {\bf F} = 0$) in the range of wavenumbers 1 $\le L_{\rm box} |{\bf k}|/2 \pi \le$ 3, such that the average forcing wavenumber was $k_f^{-1} \simeq  L_{\rm box}/2/2 \pi.$  This driving choice was made as at small scales  most of the turbulent kinetic energy is found in the solenoidal modes  \citep{Pan2010}.  

\begin{figure}
\begin{center}
\includegraphics[width=0.5\textwidth]{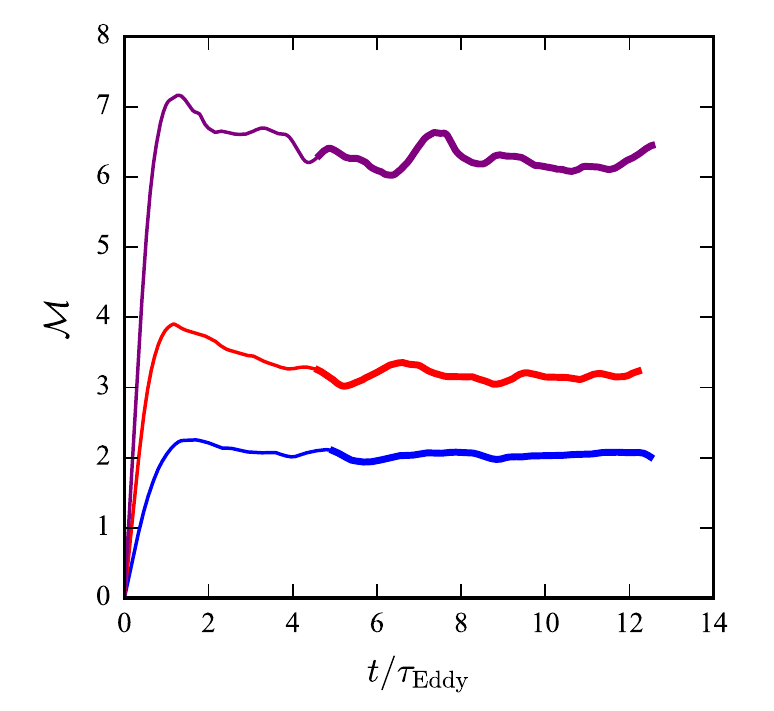}
\vspace{-0.3in}
\end{center}
\caption{Evolution of the Mach number in our numerical simulations as a function of time in units of $\tau_{\rm eddu}.$   In each run, the shaded part indicates the portion of the simulation used for our analysis, during which the average Mach numbers were 2.03, 3.17, and 6.25.
\label{Mach number}
 }
\end{figure}

To ensure a nearly-constant Mach number for each simulation, we made use of an adaptive scheme to update the driving conditions \citep{Gray2016}.    At each time step, the rms velocity was measured and the strength of the kicks was  enhanced by $\propto ({ \sigma_i}/{\sigma_t})^{-5}$, where $\sigma_i$ is the current global velocity dispersion, and $\sigma_t$ is the target value.   In this way, we carried out simulations with average Mach numbers of 2.03, 3.17, and 6.25 as shown in Figure \ref{Mach number}. 

Once the turbulence was fully developed, we assigned each cell a scalar that recorded the value of $s$ at a given time,  $t_1$. These values were then advected in a manner similar to a mass fraction
\be
\frac{\partial \rho s}{\partial t} + \nabla  \cdot \rho s {\bf v} = 0,
\ee
such that the current value of $s$ in a given cell is the mass-weighted averaged value of $s$ at the time when the passive scalar was painted on, with the weighting being computed over all the Lagrangian fluid elements that are currently in the cell.  This allows us to determine the difference between the value of a cell, $s_2,$  a time $t_2,$  and the  value of the material from which it was comprised, $s_1,$ at a previous time $t_1$.   Four such scalars were advected, which were repainted onto the simulations at various times to study the change in $s$ over a fixed number of timesteps  (250, 500, 1000, and 2000).  Defining the eddy turnover time as  $\tau_{\rm eddy} = L_{\rm box} / 2 \sigma_v$ these intervals corresponded to average time differences of
$\Delta t/\tau_{\rm eddy}$ of 0.097, 0.191, 0.383, and 0.767 in the Mach 2.03 simulation, 0.105,  0.210, 0.420, and 0.839 in the Mach 3.17 simulation, and 0.098, 0.197, 0.393, and 0.785 in the Mach 6.25 simulation.  As a convergence  test, the Mach 3 run was repeated at a resolution of $256^3$ cells. For this run, the change in $s$ and the PDF matched our $512^3$ results within 10\% over the full range of $s$ values discussed below.

\begin{figure*}[t]
\resizebox{6.5in}{!}{\includegraphics{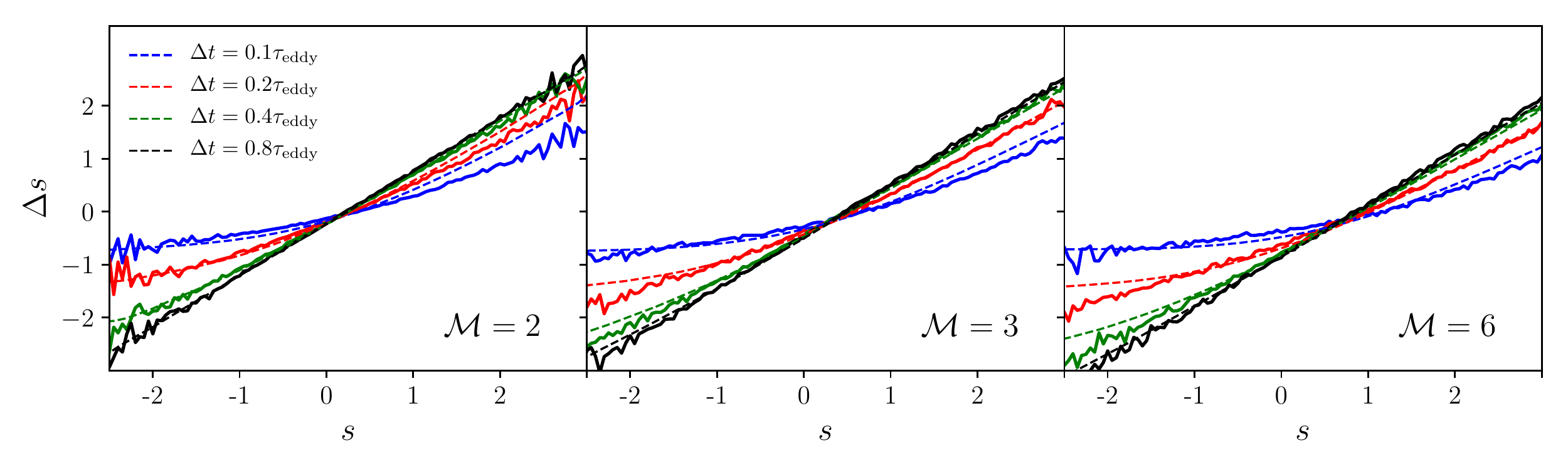}}
\resizebox{6.5in}{!}{\includegraphics{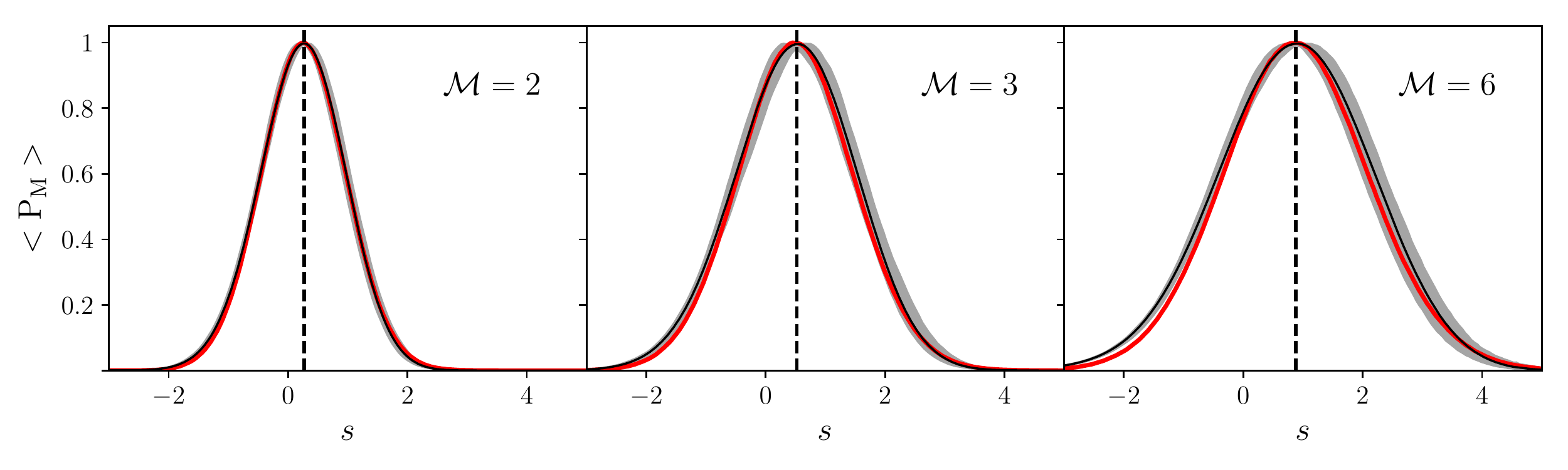}}
\vspace{-0.1in}
\caption{{\em Top:} The average value of $\Delta s$, the difference between the current value of $s$ and the mass weighted average $s$ at a previous time. The dashed lines show $\Delta s$ measured from turbulent isothermal box simulations and the solid lines are the results of Markov simulations. The panels are labeled by Mach numbers, and  the blue, red, green, and black lines correspond to  time intervals $\Delta t/\tau_{\rm eddy}$ of 0.097, 0.191, 0.383, and 0.767 in the Mach 2.03 simulation, 0.105,  0.210, 0.420, and 0.839 in the Mach 3.17 simulation, and 0.098, 0.197, 0.393, and 0.785 in the Mach 6.25 simulation.
{\em Bottom:} Mass weighted PDFs.  The solid black lines show the average value in our simulations, the shaded regions show the 16th-84th percentile range, and the red lines show the Markov results. The vertical dashed line shows the location of $s_0$, the peak of the PDF, which occurs at .27,.52, and .88.}
\label{fig:deltaspdf}
\end{figure*}

\section{Markov Model}

We then developed an analytic model of the evolution of $s$, approximating  turbulence as a temporally-homogeneous Markov process, a stochastic process for which the future state of any quantity of interest only depends on its current value.  While this is not strictly the case for a turbulent medium, in which velocity correlations are non-gaussian and correlated in time \citep{Hennebelle2012,He2017}, we found that this approximation nevertheless allowed us to accurately capture the evolution of $s$ in a simple way that provides insight into the timescales operating at low and high densities.

In this case, we can define the  propagator density function $\Pi(ds | dt, s),$ which gives the change in $s$ over a time interval $dt,$ given its initial value $s.$  Note that  $dt$ is  an infinitesimal increment, but $ds$ need not be.  However,
we can greatly simplify $\Pi(ds | d t, s)$ if we further assume that the process is continuous, meaning that $\Pi(ds | dt, s)$ varies smoothly with $s$, and is zero outside an infinitesimally small neighborhood around $ds=0$. In this case, the  propagator is a Gaussian that is completely determined by only two characterizing functions \citep{Gillespie:1996di}:  the drift or advection function, $A(s),$  and  the diffusion function, $D(s),$ as
\be
\Pi(ds | dt, s) = 
  \frac{1}{{[2 \pi D(s) dt]}^{1/2}} {\rm exp} \left\{ - \frac{[ds - A(s) dt]^2} {2 D(s) dt} \right\}.
  \label{eq:prop}
\ee
While the propagator is not measurable directly from our simulations, we can constrain its properties by comparison with a simple statistical model.
If the PDF of $s$ is exactly Gaussian, the only allowed continuous Markov process is an Ornstein-Uhlenbeck (OU) process, for which the characterizing functions take the form 
 \be
 A(s)=  [s-s_*] /\tau_{\rm ev} \qquad {\rm and} \qquad  D(s)= 2  \sigma_s^2 / \tau_{\rm ev},
 \label{eqAD}
 \ee
where  $\tau_{\rm ev}$ is a characteristic timescale for the system to evolve, and $s_*= s_0,$ and $\sigma_s^2$ set the peak and the width of the PDF as in eq.\ (\ref{eq:PDF})  \citep{Doob1942,Gillespie:1996di}.   The $\Delta s$ measured from our simulations is the change in $s$ over a fixed time interval given a final value of $s_2,$  
which in the OU case results in a linear relation:  

\be
 \Delta s (s_2, \Delta t) = [1-e^{-\Delta t/\tau_{\rm ev} }] [s_2-s_0].
 \label{eq:Deltas}
\ee
Note that this relation approaches $s_2-s_0$ at long times, which is the value expected if $s_2$ is uncorrelated with the value of log density at the beginning of the interval.

However, eq.\  (\ref{eq:Deltas}) is not a good fit to the measured value of $\Delta s$ as a function $s_2$, which is shown in the top panel of  Figure \ref{fig:deltaspdf}.  Instead, the convergence towards $\Delta s_2=s-s_0$ occurs much faster at higher values of $s_2$ than at lower values of $s_2$. In other words, material with large densities experiences changes much faster than material with lower densities, likely because the period between compressions by shocks is long compared to the timescale for the dispersal of the shocked material  \citep{Klessen2000,Vazquez2005,Glover2007,Robertson2018}.  This behavior can only be captured by moving beyond eq.\ (\ref{eqAD}).

A temporally-homogeneous  continuous Markov processes can also be expressed in terms of a Langevin equation as  \citep{gillespie1991markov}:
\be
s(t_0+dt)=s(t_0) +  A(s) dt + \mathcal{N}(0,1) \times[D(s) dt]^{1/2},
\label{eq:len}
\ee
where $\mathcal{N}(0,1)$ is a random number drawn from a normal distribution with mean of zero and variance of one. 
Note that this is a random walk as a function of time, rather than as a function of scale, as in \cite{Hopkins2013}.
While recovering an exactly Gaussian PDF  requires $\tau_{\rm ev}$ to be constant, we are able to reproduce a nearly Gaussian PDF while also reproducing the observed evolution with density
by  replacing $\tau_{\rm ev}$ with a timescale that decreases with increasing $s$, and shifting $s_*$ to preserve the position of the peak of the PDF.  This allows us to systematically measure the timescale of turbulent density changes, which we find are well modeled by the empirical formula:
\be
\tau_{\rm ev}(s) = \frac{\tau_{\rm eddy}}{3} \left[ \frac{1}{2} -\frac{1}{\pi} \arctan \left(\frac{s-s_*}{2} \right) \right],
\label{eqks}
\ee
with $s_*= \frac{3}{2} s_0.$ 
This varies slowly between $\tau_{\rm eddy}/3$ in under-dense regions and 0 in the limit of very large $s$, consistent with previous studies showing that the lifetimes of the densest regions are comparable to their (extremely short) local sound crossing times \citep[e.g.][]{Robertson2018}.

\begin{figure*}[t]
\begin{center}
\resizebox{3.4in}{!}{\includegraphics{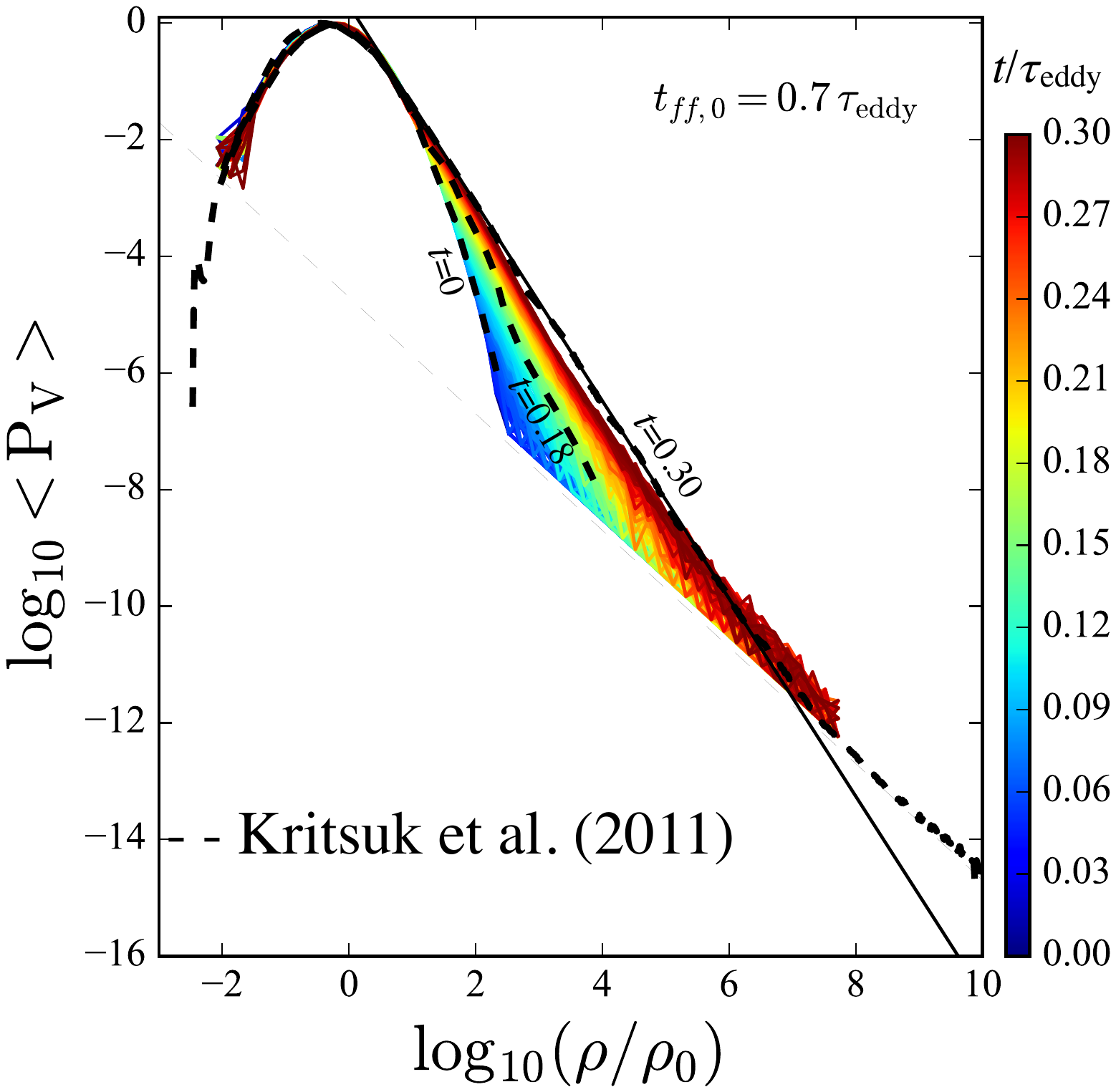}}
\resizebox{3.4in}{!}{\includegraphics{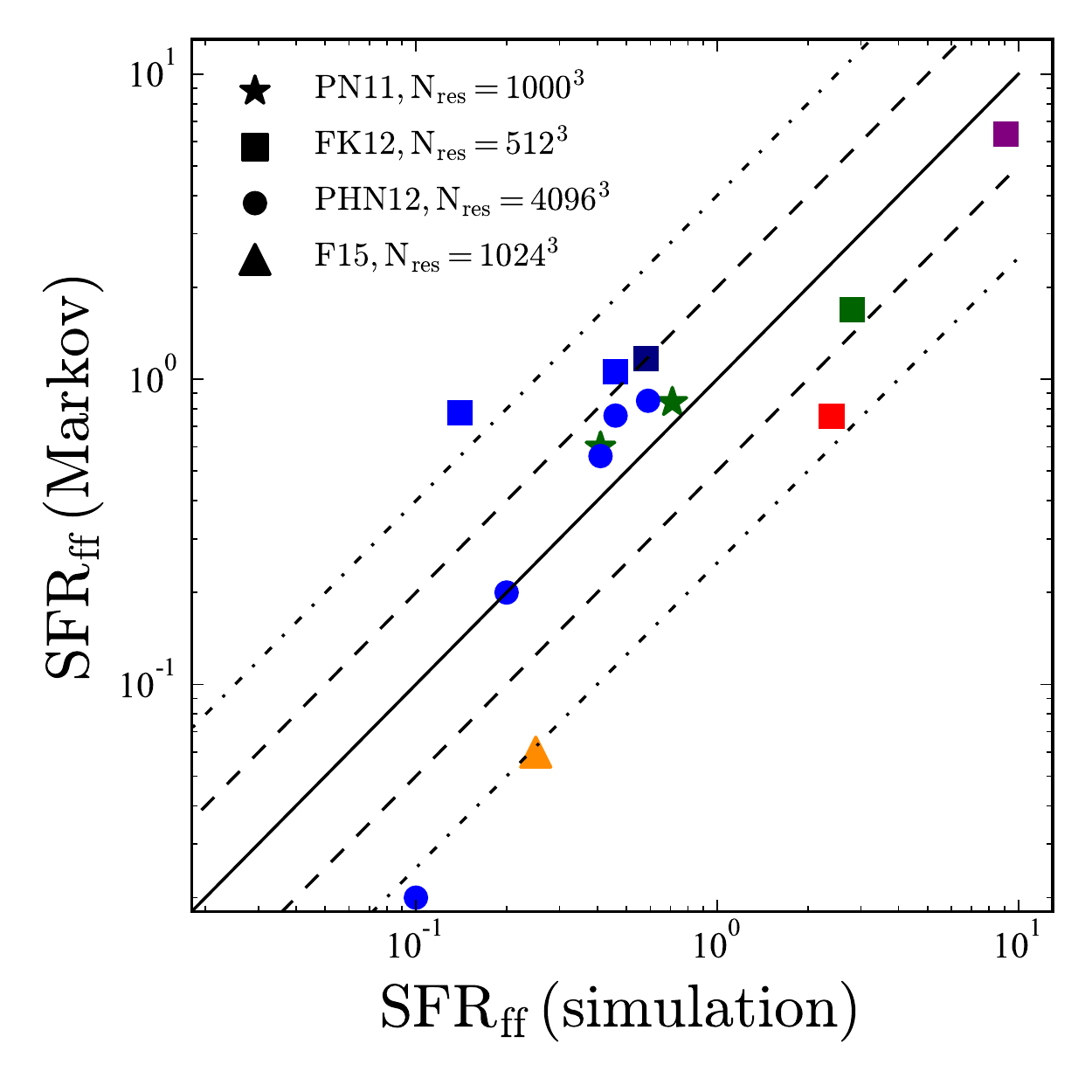}}
\end{center}
\vspace{-0.2in}
\caption{
\label{f.Kritsuk}
\emph{Left panel}: 
Comparison of the Markov results from eq.\ (\ref{e.final}) with numerical results for the evolution of the PDF of $\mathcal{M}=6$ isothermal supersonic turbulence plus gravity.
Each line represent the PDF at a given time step, color-coded as indicated. The thick black dashed lines show the numerical results  \citep{Kritsuk:2010hw} at $t=0, 0.18\tau_{\rm eddy},$ and $0.30\tau_{\rm eddy}.$ The solid black line shows a power-law with slope of $-1.695\pm0.002$ that best fits the numerical PDF for densities $\rho/\rho_0 \in [10,10^7].$ The dashed black line shows a power-law with slope of -1, that best fits the numerical results above densities $\rho/\rho_0 > 10^7$.
Note that all PDFs are volume weighted, computed as $P_V(s) \propto P_M(s)/\rho$ and renormalized.
\emph{Right panel}: Comparison of the predicted star formation rate per free fall time ($\rm SFR_{ff}$) to simulations performed by PN11 \citep{Padoan11} (stars), FK12 \citep{Federrath2012} (squares), PHN12 \citep{Padoan2012} (circles), and F15 \citep{Federrath2015} (triangle). 
The maximum resolution of each simulation is indicated in the legend. The points are color coded based on their Mach number, where
$\mathcal{M}=3,5,9,10,11$, and $50$ is shown in red, orange, green, blue, navy, and purple respectively
To match the values used in the simulations, we have imposed absorbing barriers at log densities of $s= 9$  \citep{Padoan11}, 11.5  \citep{Padoan2012}, 5.8($\mathcal{M}$=3), 6($\mathcal{M}$=9,10,11), 4.1($\mathcal{M}$=50)  \citep{Federrath2012}, and 8.75 \citep{Federrath2015}. The dashed and dot-dashed lines envelope factors of 2 and 4 from the one-to-one relation.}
 \label{fig:gravity}.
\end{figure*}

 We evolved 200,000 particles based on eqs.\  (\ref{eq:len}) and (\ref{eqks}) and used these to predict  the resulting $\Delta s (s_2,\Delta t)$ and  PDF.  These results are presented in Figure  \ref{fig:deltaspdf}, providing a good approximation to the evolution of $\Delta s$ as a function of time and Mach number.  
Using the driving routines from \cite{Federrath:2010ef}, we ran an additional $M=3.3$ hydro simulation with 50\% of the energy in compressible modes, and again found a good fit to the PDF and $\Delta s$ evolution with $s_0 = 0.7$ and eq.\ (7) unchanged. 
Thus $\tau_{\rm ev}$ is independent of $M$ and the same for mixed and solenoidal driving,
and for all cases it is significantly less than $\tau_{\rm eddy}.$    In fact, over the $s \approx s_* \pm 2$ range in which we have made our measurements $\tau_{\rm ev}(s)$ varies from $\approx \tau_{\rm eddy}/{4}$ to $\approx \tau_{\rm eddy}/{12}.$  Meanwhile, the effect of the Mach number is to increase the width of the PDF and shift its peak as required by mass conservation.

\section{Gravity and Star Formation}

Calculating the rate at which stars form requires extending the model to include gravitational collapse. 
Current analytic models do this by assigning a critical overdensity, $s_{\rm crit},$ above which gravity
overcomes turbulent motions.  In this case, the normalized star formation rate per free fall time becomes
\be
{\rm SFR_{ff}} =  \frac{ \epsilon_{\rm collapse}} {\phi_t} \int^\infty_{s_{\rm crit}} ds P_M(s) \frac{t_{\rm ff}(0)}{t_{\rm ff}(s_{\rm ff})},
\ee
where $\phi_t$ is a ``replenishment factor"  that accounts for the timescale over which the lognormal is assumed to be replenished, 
and  the efficiency factor $\epsilon_{\rm collapse}$  accounts for  the fraction of the mass with density larger than $s_{\rm crit}$ that actually collapses.
Various models in the literature make different assumptions about $s_{\rm crit}$  and take $s_{\rm ff}=$ $0$  \citep{Krumholz05},
$s_{\rm ff}=s_{\rm crit}$  \cite{Padoan11} or $s_{\rm ff}=s$  \citep{Hennebelle11,Federrath2012}, calibrating ${ \epsilon_{\rm collapse}} / {\phi_t}$ to direct numerical simulations.  Recently, \cite{Burkhart2018} developed a model in which the lognormal pdf was combined with a high-density power law tail  in a continuous and differentiable way, which has the effect of leading to higher star formation rates for lower power-law slopes, as seen in observations and simulations.

This behavior also arises naturally through a simple extension of our Markov model.
If we approximate each perturbation as a constant density sphere collapsing from rest from a large initial radius, 
conservation of energy gives $ds/dt = \left(24 \pi G \rho \right)^{1/2}.$
Adding this term to eq.\ (\ref{eq:len}), we arrive at a Langevin equation that describes supersonic, isothermal turbulence including self-gravity:
\begin{eqnarray}
s(\tilde t_0+d\tilde t)= s(t_0) +  \tilde A(s) d \tilde t + \mathcal{N}(0,1) \times \left[\tilde D(s) d \tilde t \right]^{1/2}  \nonumber \\
+ (90/\alpha)^{1/2}  \tilde L_t e^{s/2} d \tilde t,   
\label{e.final}
\end{eqnarray}
where $\tilde A\equiv \tau_{\rm eddy} A$, $\tilde D \equiv  \tau_{\rm eddy}D$ and $\tilde t \equiv  t/\tau_{\rm eddy} $ are  dimensionless such that
$\tilde A(s)$ and $\tilde D(s)$ are purely a functions  of $s,$  $\alpha \equiv 5 \sigma_v^2 R/G M$ is the virial parameter with $R$ and $M$ the cloud radius and mass, and $\tilde L_t$ is the turbulent driving scale in units of the cloud radius.
In the left panel of Figure  \ref{f.Kritsuk}, we compare the results of this modified Langevin equation with full three-dimensional gravoturbulent simulations \citep{Kritsuk:2010hw}. 
These simulations drive turbulence to a steady state with $\mathcal{M}=6$ and a corresponding  log-normal distribution with $\sigma_s =  1.73,$ and then ``turn-on" the force of gravity abruptly with $t_{\rm ff,0}=0.7~\tau_{\rm eddy}.$

To model this evolution, we start with  material distributed according to a log-normal PDF and  allow it to evolve according to eq.\ (\ref{e.final}), multiplying the gravity term by $\theta(t-t_{\rm ff,0} e^{-s/2})$ to approximate the abrupt turn-on of gravity. 
Note that our goal here is to produce a model that simply reproduces the physics included in such simulations, not to argue that such simulations capture the full physics of star formation.  The resulting evolution is shown  in the right panel of Figure \ref{f.Kritsuk}.  Here we see a power law developing  towards the high density part of the PDF whose slope matches the numerical results. 

Typically, star formation simulations convert gas into sink particles representing collapsing cloud cores once they cross a threshold value.  The normalized star formation rate per free fall time ($\rm SFR_{ff}$) is then computed as the rate at which gas is converted into sink particles divided by the mass in the simulation. In the Langevin approach, this corresponds to placing an absorbing barrier at a threshold density, and measuring the rate at which particles cross it. This method can not pick out a physical scale, but it can give an overall rate.

The right panel of Figure \ref{f.Kritsuk} shows our predicted star formation rate  compared to full simulations, with $\rm SFR_{ff}$ computed when 20\% of the gas is converted into stars. We note that different groups adopt different threshold densities, and we have adjusted our comparisons appropriately.  Furthermore, for each simulation set, we have only picked the highest resolution run. For example, we compare our results to the simulations presented in PHN12 that the have highest Alfven Mach numbers at $\mathcal{M}=10$, since we focus on pure hydro simulations. From the simulations presented in F15, we choose the pure hydro simulation without implementation of jets. 
We varied the absorbing barrier in our Markov estimate of $\epsilon_{\rm SFR}$ for the PHN12 cases, and found that higher $t_{\rm ff}/t_{\rm dyn}$ simulations converge at higher $s_{\rm abs}$ values. For example, the simulations with $t_{\rm ff}/t_{\rm dyn}$ 0.54 and 1.53 converge at $s_{\rm abs}\approx$ 8 and 11 respectively.

Although no constants have been adjusted when adding gravity to the Langevin equation, the level of agreement  in this figure is comparable to that of previous approaches after model parameters such as $s_{\rm crit}$  and  ${\epsilon_{\rm collapse}}/ {\phi_t}$  are chosen to fit the simulations.  

\section{Discussion}

The analysis here has been limited to models of star formation due to gravity acting within molecular clouds that are purely-hydrodynamic and constantly driven in near virial equillibrium. Yet, the overall method is quite general and can be applied to alternative pictures.  For example, a similar analysis could be used to quantify the evolution of the density PDF in cloud s undergoing global hierarchical collapse \citep[e.g.][]{Vazquez2017}, impacted by protostellar jets \citep[e.g.][]{Cunningham2018}, or supported by  dynamically important magnetic fields \citep[e.g.][]{Padoan2012,Federrath2015}.

The method presented here also provides a language for expressing the motion of material within pseudo-steady state systems that display a range of both densities and temperatures.  In this case, the evolution is governed by two vector functions, $\vec A(\vec s)$ and $\vec D(\vec s),$
which determine the rate at which material advects and diffuses within the two-dimensional phase space of temperature and density as
\be
 \label{eq:prop2}
\Pi(d \vec s | dt, \vec s) = 
  \frac{e^{ - \frac{[ds_1 - A_1(\vec s) dt]^2} {2 D_1(\vec s) dt} }}{{[2 \pi D_1(\vec s) dt]}^{1/2}}    
  \frac{e^{ - \frac{[ds_2 - A_2(\vec s) dt]^2} {2 D_2(\vec s) dt} }}{{[2 \pi D_2(\vec s) dt]}^{1/2}},  \\
\ee
\\
where  $\vec s \equiv (\ln \rho, \ln T).$  These functions can be measured from simulations in a similar manner to the one applied in \S3, by computing
$\Delta \vec s(\vec s_2,\Delta t),$ the change in $\vec s$ given its final value at the end of a time interval, $\Delta t$.
From these constraints, two dimensional Langevin equations can be constructed and applied to address issues such  molecule formation \cite[e.g.][]{Walch2015},
 the role of non-equillibrium processes in the ionization structure of the interstellar   and circumgalactic media \citep[e.g.][]{Richings2014,Safarzadeh2016,Gray2017,Oppenheimer2018}, and other outstanding questions throughout astrophysics.

Our work is related to generative models of log-normal distributions, in which a multiplicative process, $X_j=F_j X_{j-1},$ leads to a log-normal distribution for a random variable $X$ if $F$ is log-normally distributed. Furthermore trivial variations of a multiplicative process can generate power-law and log-normal distributions, as well as log-normal distributions with power-law tails \citep{Mitzenmacher2004}. The possibility of re-formulating our work in this way \citep[e.g.,][]{Basu:2004eq} is left for future investigation.

\acknowledgements
 We  thank Tom Abel, Marcus Br\"uggen,  Paolo Padoan, and Enrique V\'azquez-Semadeni for useful discussions, and the referee, Mark Krumholz, for his detailed comments. This work was supported by NSF grant AST14-07835 and NASA theory grant NNX15AK82G.  We  thank the Texas Advanced Computing  Center (TACC) and the Extreme Science and Engineering Discovery Environment (XSEDE) for providing HPC resources via grant  TG-AST130021. 

\vspace{-0.1in}
\bibliographystyle{apj}
%\bibliography{the_entire_lib}
%\end{document}

\expandafter\ifx\csname natexlab\endcsname\relax\def\natexlab#1{#1}\fi

\end{document}